\documentclass[reprint,amsmath,amssymb,aps,prc,floatfix]{revtex4-1}

\usepackage{graphicx}
\usepackage{dcolumn}
\usepackage{bm}
\usepackage[breaklinks=true]{hyperref}
\usepackage{xcolor}
\usepackage{ulem}

\begin{document}
\title{Microscopic effective reaction theory for deuteron-induced reactions}
\author{Yuen Sim Neoh}
  \email{neohys@rcnp.osaka-u.ac.jp}
\author{Kazuki Yoshida}
\author{Kosho Minomo}
\author{Kazuyuki Ogata}
\affiliation{%
 Research Center for Nuclear Physics, Osaka University, Ibaraki 567-0047, Japan
}%
\date{\today}
\begin{abstract}
The microscopic effective reaction theory is applied to deuteron-induced reactions. A reaction model-space characterized by a $p+n+{\rm A}$ three-body model is adopted, where A is the target nucleus, and the nucleon-target potential is described by a microscopic folding model based on an effective nucleon-nucleon interaction in nuclear medium and a one-body nuclear density of A. The three-body scattering wave function in the model space is obtained with the continuum-discretized coupled-channels method (CDCC), and the eikonal reaction theory (ERT), an extension of CDCC, is applied to the calculation of neutron removal cross sections. Elastic scattering cross sections of deuteron on $^{58}$Ni and $^{208}$Pb target nuclei at several energies are compared with experimental data. The total reaction cross sections and the neutron removal cross sections at 56 MeV on 14 target nuclei are calculated and compared with experimental values.\\\\

\begin{description}
\item[PACS numbers]
24.10.Eq, 25.45.-z
\end{description}
\end{abstract}

\maketitle


\section{\label{sec1} Introduction}

Projectile-breakup reactions have been utilized for studying the structures of nuclei, in the field of physics of unstable nuclei in particular. One of the most successful models for describing breakup processes of a nucleus into a few fragments is the continuum-discretized coupled-channels method (CDCC)~\citep{Kam86,AUSTERN1987125,Yahiro01012012}. The theoretical foundation of CDCC through its relation to the distorted-wave Faddeev theory was given in Refs.~\cite{Aus89,Aus96}. Recently, progress has been made to construct scattering potentials employed in CDCC calculation from a microscopic point of view. This attempt can be interpreted as an application of the microscopic reaction theory~\citep{Yahiro01012012}, which is based on the multiple scattering theory for nucleus-nucleus scattering~\citep{Yahiro01102008}, to reaction processes involving a weakly-bound projectile, by setting an appropriate model space. Very recently, a microscopic description of the projectile wave function also has been reported~\citep{Descouvemont2015}. Such a  microscopic description of breakup processes is crucial for quantitative determination of structure of unstable nuclei from reaction observables.

In this work, we discuss the breakup of deuteron, the simplest composite nucleus, to examine the microscopic effective reaction theory, i.e., CDCC with an appropriate reaction model-space and the microscopic nucleon-nucleus ($N$-A) optical potentials. Nowadays, $N$-A scattering observables can be described microscopically with no free adjustable parameter~\cite{Amo00,Toy14,Toy15}. It is thus expected that the microscopic CDCC calculation can describe deuteron elastic scattering data that have successfully been reproduced by CDCC with phenomenological $N$-A optical potentials~\cite{Kam86}. It should be pointed out that the model in the present study differs from a fully microscopic calculation such as those of \textit{ab-initio} method
\citep{PhysRevC.83.044609}. The limitation of the three-body model description
compared with a many-body reaction theory can be seen in, e.g.,
Ref.~\citep{PhysRevC.89.024605}. Nevertheless, the deuteron elastic scattering and
its elastic breakup processes have successfully been described by CDCC,
a three-body reaction model, with phenomenological optical potentials.
The purpose of the present study is to implement microscopic optical potentials
in the usual CDCC calculation. This can be regarded as an \textit{effective}
microscopic reaction model based on a three-body model.

It should be noted that the deuteron breakup states in the intermediate channel, i.e., the virtual breakup of deuteron, was shown to be crucial for describing the deuteron elastic scattering \citep{Kam86,AUSTERN1987125}. Convergence of the CDCC results with respect to the $p$-$n$ relative orbital angular momentum $\ell$ (including odd partial waves) is discussed as well as the Coulomb breakup contribution. Furthermore, we apply the eikonal reaction theory (ERT)~\citep{Yahiro2011} with microscopic optical potentials, microscopic ERT, to the calculation of the total reaction cross sections $\sigma_{\rm R}$ and the neutron removal cross sections $\sigma_{-n}$ on various target nuclei measured at 28~MeV/nucleon~\citep{MATSUOKA19801}. ERT is an extension of CDCC for inclusive breakup observables and can be interpreted as also the extension of the Glauber model to explicitly take into account the $p$-A Coulomb interaction, hence the Coulomb breakup effects on those observables.

 In Sec.~\ref{sec2.0}, we describe the method of microscopic CDCC, followed by the methodology used to extract inclusive neutron removal cross section using ERT. The numerical results of microscopic CDCC and ERT are presented in Sec.~\ref{sec3.0}. In Sec.~\ref{sec4.0} we give a summary.
\section{\label{sec2.0}Framework}
\begin{figure}[ht]
\includegraphics{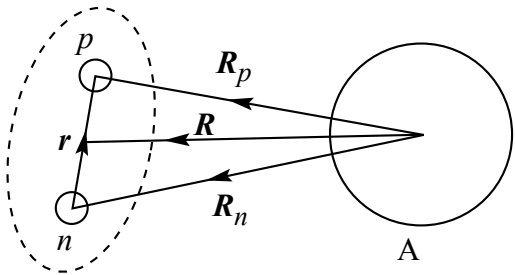}
\caption{\label{cdcc_schematic}Schematics of deuteron-induced nuclear reaction system.}
\end{figure}%
As deuteron is well-known to be a weakly bound nucleus consisting of proton (\textit{p}) and neutron (\textit{n}), deuteron-induced nuclear reactions are, as a result of an appropriate choice of the reaction model-space, often treated as a $p+n+{\rm A}$ three-body system. The total wave function $\Psi$ satisfies the Schr\"{o}dinger equation
\begin{equation}\label{se}
  \left(H-E\right)\Psi\left(\bm{R},\bm{r}\right)=0,
\end{equation}
where the coordinates $\bm{R}$ and $\bm{r}$ are defined as shown in Fig.~\ref{cdcc_schematic} and the total Hamiltonian $H$ is the sum of the relative kinetic operators and the potentials in the center-of-mass frame. In the three-body system, the total potential is written as $V_{pn}(\bm{r})+U_p(\bm{R}_p)+V_{\text{C}}(\bm{R}_p)+U_n(\bm{R}_n)$; $V_{pn}$ is the interaction between $p$ and $n$, $U_p$ $(U_n)$ is the nuclear potential between $p$ $(n)$ and A, and $V_{\rm C}$ is the Coulomb potential between $p$ and A. A general overview of the microscopic construction of potentials $U_p$ and $U_n$ is given in Ref.~\citep{Yahiro01012012}. According to that, in the present work, we adopt the microscopic nucleon-target potentials obtained by folding the Melbourne \textit{g}-matrix interaction \citep{Amo00} with a target density in the same manner as in Ref.~\cite{Toyokawa2013}. Target densities used in the folding procedure are obtained by the Hartree-Fock calculation with the Gogny D1S force
as explained in Ref.~\citep{0954-3899-37-8-085011}. The validity of the microscopic optical potentials for proton elastic scattering is shown in Refs.~\cite{Toyokawa2013} and \citep{0954-3899-37-8-085011}. Then Eq.~\ref{se} with the microscopic optical potentials is solved by CDCC~\citep{AUSTERN1987125,Yahiro01012012,Kam86}.

In our microscopic optical potential the nonlocality coming from knockon-exchange process is taken into account
properly with the localization method by Brieva and Rook  \citep{0954-3899-37-8-085011} but the potential is still energy-dependent.
We adopt the choice of $E_d/2$ as the energy of optical potentials in the present study. This prescription is valid for elastic scattering as discussed in Refs.~\citep{JOHNSON1972619, PhysRevLett.110.112501}.
For deuteron stripping reaction $(d,p)$, Timofeyuk and Johnson suggested that a different choice of the energy of
optical potentials is needed \citep{PhysRevLett.110.112501}, as mentioned above.
It is nontrivial whether the energy-dependence of optical potential is crucial or not for other reaction observables.
However, we can expect that the energy-dependence coming from a nonlocal effect is not essential for elastic breakup
and neutron removal reactions since such reactions are determined by the asymptotic behavior of scattering wave functions.

We denote the neutron removal cross section from deuteron as $\sigma_{-n}$. It can be written as the sum
\begin{equation}\label{total_nucleon_removal}
  \sigma_{-n} = \sigma_{\text{EB}} + \sigma_{n:\text{STR}},
\end{equation}
where $\sigma_{\text{EB}}$ is the elastic breakup cross section and $\sigma_{n:\text{STR}}$ is the neutron stripping cross section; in the elastic breakup A stays in the ground state, whereas in the stripping A is excited to unspecified states. $\sigma_{\text{EB}}$ is calculated with CDCC and $\sigma_{n:\text{STR}}$ with ERT~\citep{Yahiro2011,Hashimoto2011}. The essence of ERT is that the scattering matrix is divided into the proton and neutron parts, $S_p$ and $S_n$, respectively, by using the adiabatic approximation to only $U_n$ \citep{Hashimoto2011}. With $S_p$ and $S_n$ thus separated, we can define $\sigma_{n:\text{STR}}$ as
\begin{equation}\label{n_str}
  \sigma_{n:\text{STR}} = 2\pi\int \langle 0\lvert \lvert S_p\rvert ^2 \left( 1-\lvert S_n\rvert ^2 \right) \rvert 0\rangle bdb,
\end{equation}
where $b$ is the impact parameter and $\lvert 0\rangle$ is the ground state of deuteron. Equation~(\ref{n_str}) has the simple interpretation regarding projectile deuteron; it is the joint probability of the proton being scattered off (survival probability) and the neutron being absorbed out of the model space (absorption probability) simultaneously. From Eq.~(\ref{n_str}), one sees \citep{Hashimoto2011} that
\begin{equation}
  \sigma_{n:\text{STR}} = \sigma_\text{R}-\sigma_{\text{EB}} - \left(\sigma_{\text{R}}(p)-\sigma_{\text{EB}}(p)\right),
\end{equation}
where $\sigma_{\text{R}}(p)$ and $\sigma_{\text{EB}}(p)$ are the total reaction and elastic breakup cross sections, respectively, obtained by solving Eq.~(\ref{se}) without $U_n$. Thus Eq.~(\ref{total_nucleon_removal}) becomes
\begin{equation}
  \sigma_{-n} = \sigma_{\text{R}} - \sigma_{\text{R}}(p) + \sigma_{\text{EB}}(p).\label{sigma-n}
\end{equation}
Equation~(\ref{sigma-n}) gives the prescription formula to obtain the inclusive neutron removal cross section from deuteron, whose components can be obtained with CDCC.
\section{\label{sec3.0} Results and Discussion}
In the description of the $p$-$n$ system, we adopt a one-range Gaussian interaction \citep{Ohmura01021970} as $V_{pn}$. The intrinsic spin \textit{S} of deuteron is assumed to be zero for simplicity; the effect of such non-zero intrinsic spin will be discussed later for high-energy case. The \textit{p}-\textit{n} breakup states with s-, p-, d-, f-, g-, h-, and i-partial waves are included. Each of them is discretized by the momentum-bin method with an equal increment $\Delta k =0.1 $ fm$^{-1}$ to a maximum of $k_{\text{max}}=1.0\text{ fm}^{-1}$. The maximum values of $r$ and $R$ are 100 fm and 250 fm, respectively. 
\begin{figure}
\includegraphics{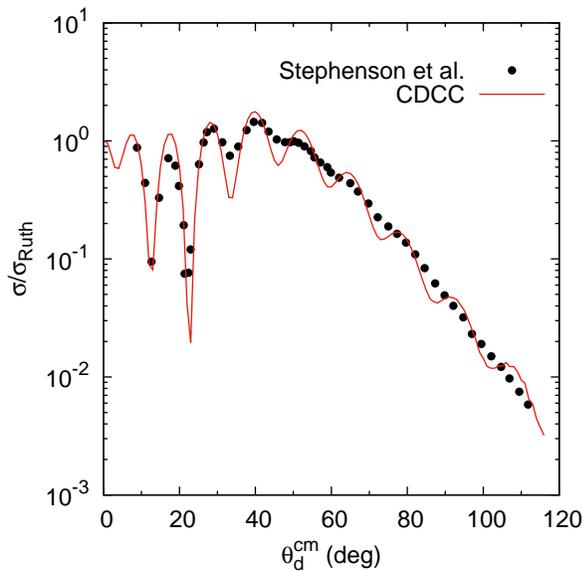}
\caption{\label{80MeV}Deuteron elastic scattering differential cross section normalized to the Rutherford cross section at 80 MeV on $^{58}$Ni. The solid line denotes the result of the microscopic CDCC calculation. The experimental data are taken from Ref.~\citep{Stephenson1981316}.}
\end{figure}

\subsection{\label{sec3.1}Deuteron Elastic Scattering}
\begin{figure}
\includegraphics{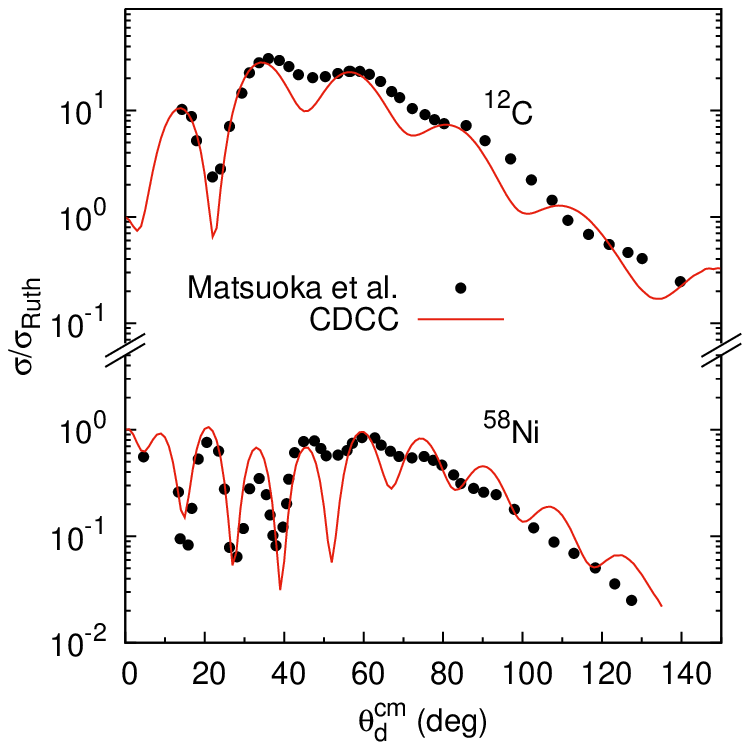}
\caption{\label{56MeV}Deuteron elastic scattering differential cross section normalized to the Rutherford cross section at 56 MeV on $^{12}$C and $^{58}$Ni. The solid lines denote the results of the microscopic CDCC calculation. The experimental data are taken from Ref.~\citep{MATSUOKA1986413}.}

\includegraphics{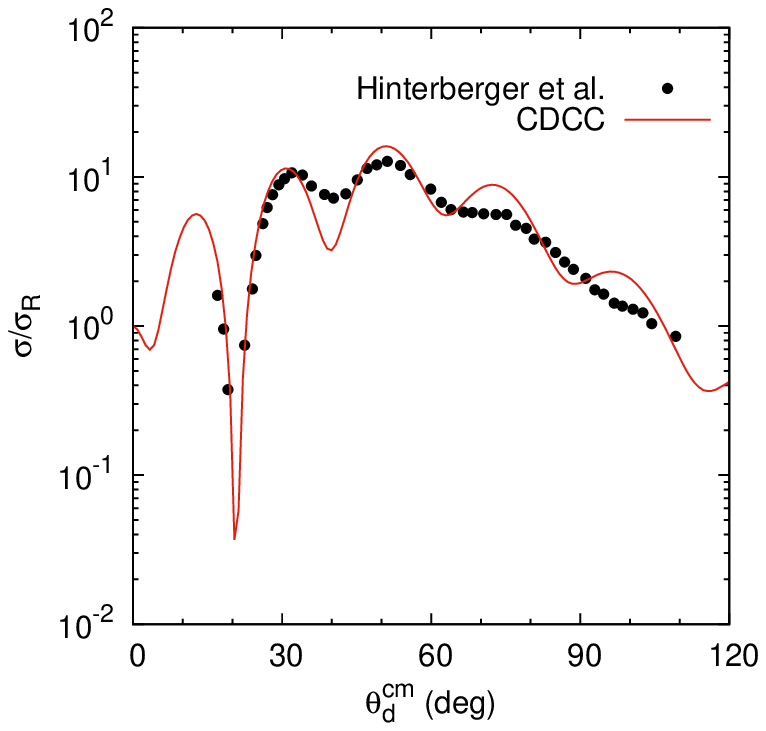}
\caption{\label{52MeV}Deuteron elastic scattering differential cross section normalized to the Rutherford cross section at 52 MeV on $^{16}$O. The solid line denotes the result of the microscopic CDCC calculation. The experimental data are taken from Ref.~\citep{HINTERBERGER1968265}.}
\end{figure}

Figure~\ref{80MeV} shows the elastic differential cross sections of deuteron scattering on $^{58}$Ni at 80 MeV. Figure~\ref{56MeV} shows the elastic scattering result of 56 MeV deuteron on $^{12}$C and $^{58}$Ni, while Fig.~\ref{52MeV} shows the elastic scattering of 52 MeV deuteron on $^{16}$O. The microscopic calculations show good agreement with the experimental data~\cite{Stephenson1981316} up to the large angles without any free parameters. At these energies, the microscopic CDCC calculation shows better agreement than the previous calculation \citep{Yahiro01041986} with the phenomenological potential, even though the tensor coupling between the deuteron s- and d-states is not included here.

\begin{figure}
\includegraphics{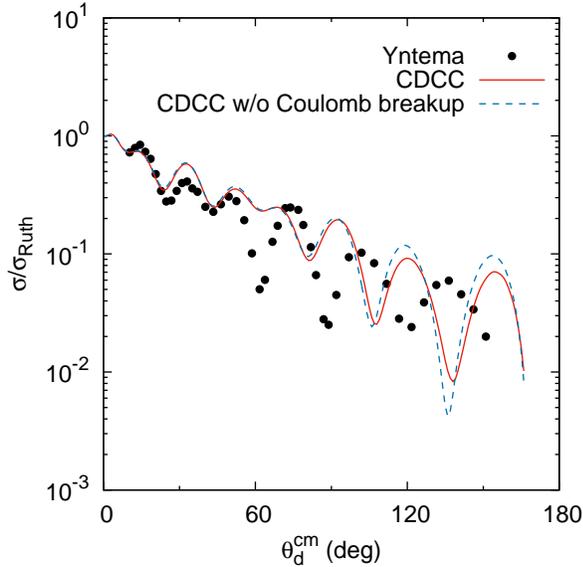}
\caption{\label{d58Ni21.6MeV}Same as Fig.~\ref{80MeV} but at 21.6 MeV. The dashed line corresponds to the result of CDCC without the Coulomb breakup. The experimental data are taken from Ref.~\citep{PhysRev.113.261}.}
\end{figure}

However, the microscopic calculation (solid line) at 21.6 MeV cannot reproduce the data at backward angles (Fig.~\ref{d58Ni21.6MeV}). This may indicate that the microscopic optical potentials are invalid at such low energies (around 10 MeV for nucleon energy). This will be reasonable because in the folding model calculation adopted, only the knockon exchange is taken into account. In other words, the total wave function of the nucleon-nucleus system is not fully antisymmetrized. As discussed in Ref.~\citep{PhysRev.97.1336}, at higher energies this treatment is justified. At low energies, however, other complicated exchange process will become important and the present folding model procedure will lose the justification.

The advantage of the present calculation compared to previous studies is twofold. First, we include not only even- but also odd-partial $\ell$ waves of the $p$-$n$ breakup states which are previously ignored in Refs.~\citep{Yahiro01041986,Piyadasa1989,Piyadasa1999}. The odd-partial waves contribute mainly to the Coulomb breakup processes that are expected to be important at lower incident energies. Secondly, we clarify the role of $\ell>2$ to the elastic and breakup cross section which is not taken into account by previous studies such as those in Ref.~\citep{CHAUHUUTAI200656,CHAUHUUTAI200680}. In particular, we can confirm that $\ell_{\text{max}}=2$ is sufficiently convergent for elastic scattering cross section but not for breakup observables, which will be discussed in more detail in Sec.~\ref{sec3.2}.

We show by the dashed line in Fig.~\ref{d58Ni21.6MeV} the result of microscopic CDCC without Coulomb breakup. One sees that the difference is very small except at backward angles. This is also confirmed in the case on $^{208}$Pb at 56 MeV,
as shown in Fig.~\ref{d208Pb56MeV} below.
Therefore, we can conclude that the Coulomb breakup effects are negligibly small
for deuteron elastic scattering, which validates the previous studies \citep{Yahiro01041986,Piyadasa1989,Piyadasa1999,CHAUHUUTAI200656,CHAUHUUTAI200680}. It should be noted, however,  that the Coulomb breakup effects are significant for breakup cross sections
as shown in Sec.~\ref{sec3.2}.

\begin{figure}
\includegraphics{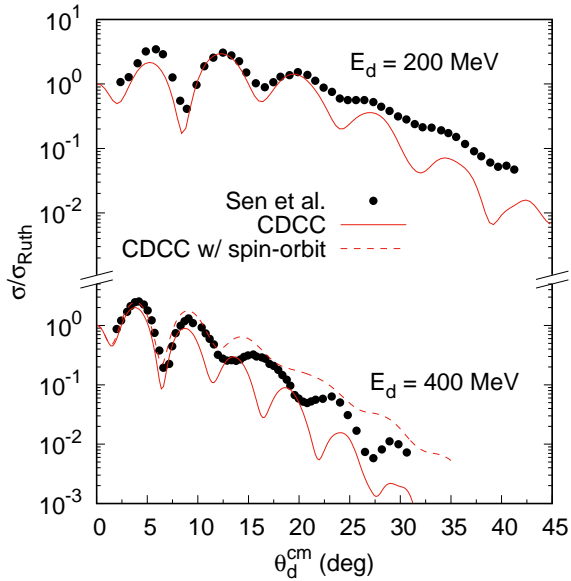}
\caption{\label{d58Ni200_400MeV}Same as Fig.~\ref{80MeV} but at 200 MeV and 400 MeV. The dashed line is the result of microscopic CDCC with the spin-orbit interaction. The experimental data are taken from Ref.~\citep{VanSen1985185}.}
\end{figure}
\begin{figure}
\includegraphics{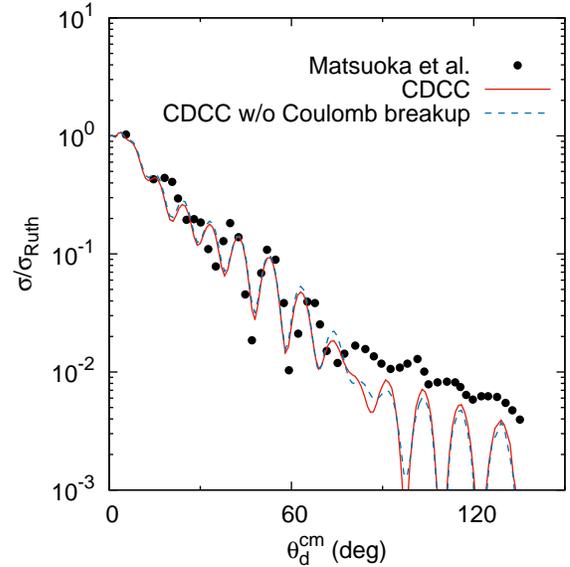}
\caption{\label{d208Pb56MeV}Same as Fig.~\ref{d58Ni21.6MeV} but at 56 MeV on $^{208}$Pb. The experimental data are taken from Ref.~\citep{MATSUOKA1986413}.}
\end{figure}

Figure~\ref{d58Ni200_400MeV} shows the elastic scattering cross sections of deuteron at higher energies, 200 and 400 MeV, on $^{58}$Ni.
The microscopic CDCC calculation (solid line) can reproduce the data fairly well at 200 MeV but it undershoots the data at 400 MeV. As mentioned above, the spin-orbit part of the microscopic optical potential has been ignored in the present study.
However, it is well known that at hundreds of MeV/nucleon, the effect of the spin-orbit potential
becomes more apparent. Therefore, at 400 MeV, we perform CDCC calculation including the spin-orbit potential; we assume $S=1$ and include the s-, p-, and d-wave breakup states. The result is shown by the dashed line in Fig.~\ref{d58Ni200_400MeV}. One can find that the effect of the spin-orbit potential gives good agreement between the calculated and measured cross sections, at forward angles in particular. We have confirmed that the effect of the spin-orbit interaction is small in other reactions discussed in the present study.

Figure~\ref{d208Pb56MeV} shows the result at 56~MeV on $^{208}$Pb. The microscopic CDCC is quite successful in reproducing the experimental data, as in Fig.~\ref{56MeV}. Even with the $^{208}$Pb target, as mentioned above, the contribution of the Coulomb breakup is very small on the elastic cross section.

\begin{figure}
\includegraphics{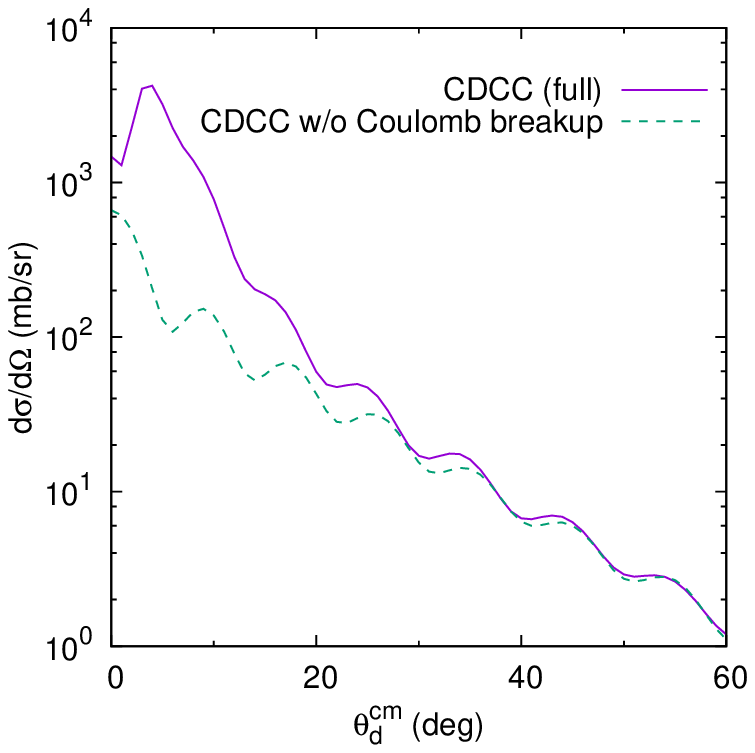}
\caption{\label{breakup}Angular distribution of the total elastic breakup cross section at 56 MeV on $^{208}$Pb. The solid line denotes the result of full CDCC, whereas the dashed line denotes CDCC without Coulomb breakup.}
\end{figure}
 
\subsection{\label{sec3.2}Breakup Observables}
\begin{table*}
\caption{\label{table1}Decomposition of elastic breakup cross section into partial wave contributions from s-wave to i-wave. All cross sections are in units of mb.}
\begin{ruledtabular}
\begin{tabular}{cccccccccc}
 Target & Energy (MeV) & $\sigma_{\text{EB}}$ & s & p & d & f & g & h & i\\ \hline
 $^{58}$Ni & 21.6 & 72.82 & 10.27 & 19.78 & 17.78 & 5.56 & 7.56 & 2.67 & 9.20 \\
 $^{58}$Ni & 56 & 149.01 & 24.76 & 42.96 & 39.28 & 4.75 & 17.92 & 2.12 & 17.22 \\
 $^{58}$Ni & 80 & 157.54 & 27.32 & 48.36 & 51.72 & 5.42 & 15.61 & 1.95 & 7.15 \\
 $^{58}$Ni & 200 & 111.73 & 19.35 & 46.95 & 38.74 & 2.34 & 3.51 & 0.26 & 0.59 \\
 $^{58}$Ni & 400 & 53.21 & 6.89 & 31.87 & 13.25 & 0.30 & 0.76 & 0.04 & 0.10 \\
 $^{12}$C & 56 & 105.51 & 22.48 & 4.14 & 57.91 & 0.83 & 14.31 & 0.31 & 5.52 \\
 $^{25}$Mg & 56 & 121.29 & 24.47 & 11.95 & 52.55 & 1.95 & 18.72 & 0.76 & 10.89\\
 $^{27}$Al & 56 & 115.38 & 22.66 & 13.55 & 48.93 & 2.12 & 17.15 & 0.81 & 10.15\\
 $^{48}$Ti & 56 & 137.84 & 25.21 & 29.56 & 42.67 & 3.45 & 19.06 & 1.54 & 16.35\\
 $^{51}$V & 56 & 139.88 & 25.31 & 31.29 & 41.65 & 3.65 & 19.41 & 1.61 & 16.96\\
 $^{54}$Fe & 56 & 146.45 & 24.93 & 38.64 & 40.22 & 4.54 & 18.93 & 1.98 & 17.20 \\
 $^{89}$Y & 56 & 170.15 & 27.14 & 64.12 & 37.40 & 6.53 & 14.91 & 2.78 & 17.27 \\
 $^{90}$Zr & 56 & 172.69 & 27.27 & 66.62 & 37.44 & 6.77 & 14.68 & 2.83 & 17.08 \\
 $^{118}$Sn & 56 & 191.22 & 29.52 & 84.69 & 37.59 & 9.57 & 12.25 & 3.47 & 14.12 \\
 $^{159}$Tb & 56 & 231.43 & 36.51 & 111.69 & 43.42 & 14.15 & 10.47 & 4.31 & 10.88\\
 $^{181}$Ta & 56 & 257.84 & 41.71 & 125.48 & 48.41 & 17.00 & 10.44 & 4.86 & 9.94 \\
 $^{197}$Au & 56 & 277.90 & 45.78 & 135.73 & 52.63 & 19.15 & 10.35 & 5.22 & 9.04\\
 $^{208}$Pb & 56 & 287.50 & 49.47 & 139.75 & 55.38 & 18.65 & 10.40 & 5.08 & 8.77 \\
 $^{209}$Bi & 56 & 288.93 & 48.01 & 141.80 & 55.05 & 20.36 & 10.10 & 5.41 & 8.20 \\
\end{tabular}
\end{ruledtabular}
\end{table*}

\begin{table*}
\caption{\label{table2}Total reaction cross sections and neutron removal cross sections of 56 MeV deuteron on 14 target elements. Experimental data are taken from Ref.~\citep{MATSUOKA19801}. All values shown are in units of mb.}
\begin{ruledtabular}
\begin{tabular}{ccccc}
 &\multicolumn{2}{c}{Total reaction cross section, $\sigma_{\text{R}}$}&\multicolumn{2}{c}{Neutron removal cross section, $\sigma_{-n}$}\\
 Target& CDCC & data &  CDCC & data \\ \hline
 $^{12}$C& 832 & 876 & 347 & 209 \\
 $^{25}$Mg& 1182 & 1162 & 422 & 299 \\
 $^{27}$Al& 1234 & 1198 & 432 & 309 \\
 $^{48}$Ti& 1577 & 1528 & 488 & 458 \\
 $^{51}$V& 1609 & 1570 & 491 & 481 \\
 $^{54}$Fe& 1637 & 1610 & 519 & 550 \\
 $^{58}$Ni& 1692 & 1662 & 524 &582 \\
 $^{89}$Y& 2055 & 2021 & 580 & 661 \\
 $^{90}$Zr& 2065 &2032 & 590 & 705 \\
 $^{118}$Sn& 2349 &2310 & 620 & 735 \\
 $^{159}$Tb& 2680 &2671 & 708 & 847 \\
 $^{181}$Ta& 2791 &2848 & 761 & 1011 \\
 $^{197}$Au& 2881 &2972 & 799 & 1055 \\
 $^{209}$Bi& 2978 &3061 & 810 & 1082 \\
\end{tabular}
\end{ruledtabular}
\end{table*}
In Table~\ref{table1} we show the total elastic breakup cross section $\sigma_{\rm EB}$ and its decomposition into the contributions $\sigma_{{\rm EB};\ell}$ from $p$-$n$ partial waves for each reaction system discussed in Sec.~\ref{sec3.1} and below. As mentioned, the odd waves contribute to $\sigma_{\rm EB}$ mainly through the Coulomb breakup. Among them, the p-wave contribution $\sigma_{{\rm EB};1}$ is dominant because of the properties of the multipoles of the $p$-A Coulomb interaction. The contributions from $\ell=3$ and 5 are much smaller than $\sigma_{{\rm EB};1}$, showing a clear convergence of the result with respect to odd $\ell$. On the other hand, for even partial waves, the convergence is very slow at 56~MeV and lower energies. At 21.6~MeV on $^{58}$Ni, $\sigma_{{\rm EB};6}$ is even larger than $\sigma_{{\rm EB};4}$. Nevertheless, fortunately, it is found that adding $\ell=8$ gives a very small change in $\sigma_{\rm EB}$; typically it is less than about 2\% and, at maximum, it is 4.5\% for the reaction at 56~MeV on $^{208}$Pb. Thus we conclude that our CDCC calculation converges at $\ell_{\rm max}=6$ within a few percent error. To determine the role of higher partial waves $\ell>2$ on elastic breakup cross section, we found that at 56 MeV on $^{208}$Pb, $\sigma_{\text{EB}}$ at $\ell_{\text{max}}=2$ differs with $\ell_{\text{max}}=6$ by about 8 \%, while at 21.6 MeV on $^{58}$Ni, their difference becomes larger at about 26 \%. We conclude that elastic breakup cross section does not sufficiently converge at $\ell_{\text{max}}=2$, especially at lower energies.

At higher energies, one can directly see a clear convergence with $\ell$ from the values shown in Table~\ref{table1}.
Another important finding is that the Coulomb breakup plays an important role also for the light targets. In fact, the Coulomb breakup contributes as much as 18\% (30\%) to $\sigma_{{\rm EB}}$ at 56~MeV on $^{12}$C ($^{58}$Ni). To see the role of the Coulomb breakup in more detail, we show in Fig.~\ref{breakup} the angular distribution of the total elastic breakup cross section at 56 MeV on $^{208}$Pb. The solid line is the result of the full CDCC calculation and the dashed line is that without \textit{p}-A Coulomb interaction. As expected, the Coulomb breakup is dominant  at forward angles; neglecting the Coulomb breakup decreases the breakup cross section by an order of magnitude at around $5^\circ$.
\begin{figure}
\includegraphics{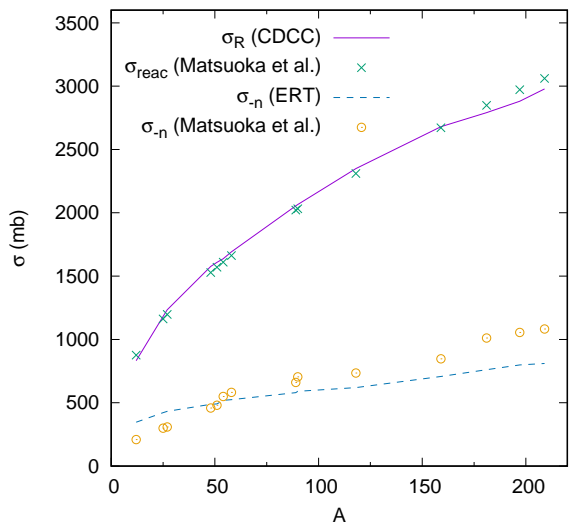}
\caption{\label{-n} Comparisons of microscopic CDCC calculation of total reaction cross sections (solid) and neutron removal cross sections (dashed) with the experimental data \citep{MATSUOKA19801}.}
\end{figure}
Finally, we show in Table~\ref{table2} and Fig.~\ref{-n} the results of the total reaction cross sections $\sigma_{\rm R}$ and the neutron removal cross sections $\sigma_{-n}$ of 56~MeV deuteron on 14 target nuclei. One can see that the results agree fairly well with experimental data, for $\sigma_{\rm R}$ in particular.

\section{\label{sec4.0}Summary}
We have carried out CDCC calculation by employing microscopic scattering potentials, and showed that the observables obtainable from this method agree fairly well with experimental data in a wide range of energies without the need for artificial normalization or any arbitrary parameters. The Coulomb breakup is explicitly taken into account and its contribution to the total
elastic breakup cross section $\sigma_{\text{EB}}$ turned out to be significant at forward angles in the case of 56 MeV incident deuteron energy on $^{208}$Pb. Furthermore, its contribution to $\sigma_{\text{EB}}$
was shown to be somewhat large even for light target nuclei. On the other hand, the Coulomb breakup effect on the elastic cross section was found to be negligible in all the cases investigated in the present study. The total reaction cross sections and the inclusive nucleon removal cross sections provided by ERT also agrees fairly well with the experimental data, demonstrating the possible use-case for other nucleon-removal cross section studies in future.

In conclusion, the microscopic CDCC, which is a microscopic effective reaction theory for deuteron-induced reactions, has been shown to be able to describe the existing experimental data for many reaction systems with no free parameters. The success of the microscopic CDCC will be of great importance for proceeding with further studies on deuteron-induced reactions in which unstable nuclei are involved.

\begin{acknowledgments}
The authors are thankful to T. Matsumoto and S. Nakayama for the valuable discussions. This work was supported in part by Grants-in-Aid of the Japan Society
for the Promotion of Science (Grants No. JP15J01392, No. JP16K17698 and JPT16K053520) and by the ImPACT Program of the Council for Science, Technology and
Innovation (Cabinet Office, Government of Japan).
\end{acknowledgments}
\bibliography{ref}

\end{document}